\documentclass[a4paper,11pt]{article}
\pdfoutput=1 

\usepackage{jcappub} 

\usepackage[T1]{fontenc} 
\usepackage{graphicx}

\def\apj{ApJ}

\def\mnras{MNRAS}

\def\apjl{ApJ Letters}

\title {\boldmath The roles of environment and interactions on the evolution of red and blue galaxies in the EAGLE simulation}

\author[a]{Apashanka Das} 

\author[a]{and Biswajit Pandey}

 \affiliation[a]{Department of Physics, Visva-Bharati University,
  Santiniketan, 731235, India}
\emailAdd{a.das.cosmo@gmail.com}
\emailAdd{biswap@visva-bharati.ac.in}

\abstract{We study the evolution of the red and blue galaxies from
  $z=3$ to $z=0$ using the EAGLE simulation. The galaxies in the blue
  cloud and the red sequence are separated at each redshift using a
  scheme based on Otsu's method. Our analysis shows that the two
  populations have small differences in the local density and the
  clustering strength until $z=2$, after which the red galaxies
  preferentially occupy the denser regions and exhibit a significantly
  stronger clustering than the blue galaxies. The significant
  disparities in cold gas mass and specific star formation rate (sSFR)
  observed before $z=2$ suggest that factors beyond environmental
  influences may also contribute to the observed
  dichotomy. Interacting galaxy pairs at a given separation exhibit a
  higher SFR at increasing redshifts, which may be linked to the
  rising gas fractions at higher redshift. As redshift decreases, the
  SFR decreases across all separations, suggesting a gradual depletion
  of the cold gas reservoir. At pair separations $<50$ kpc, an
  anomalous increase in the SFR among paired galaxies in isolation
  around $z \sim 2$ suggests that environmental effects begin to
  dominate at this redshift, thereby increasing the rate of galaxy
  interactions and the occurrence of starburst galaxies. We observe a
  substantial decrease in the blue fraction in paired galaxies
  starting from $z=1$ to the present. However, the decrease in the
  blue fraction in paired galaxies with their second nearest neighbour
  at a distance greater than 500 kpc continues until $z=0.5$, after
  which the blue fraction begins to increase.}
        
\begin{document}
\maketitle
\flushbottom


\section{Introduction}

The observed bimodality \citep{strateva, blanton03, bell1, balogh,
  baldry04a} in the colour distribution of galaxies indicates that
they fall into two distinct populations: a blue, star-forming
population, and a red, quiescent population. This bimodality is
indicative of a dichotomy in the star formation histories of galaxies,
supporting the idea of a transformation from blue to red over time.

The blue galaxies dominate the cosmic landscape at earlier
times. These galaxies are characterized by a high star formation rate,
with young, hot, and massive stars contributing to their blue
colour. The prevalence of the blue galaxies at earlier times suggests
a period of intense star formation fueled by abundant gas reservoirs
and galactic interactions. These interactions trigger gravitational
instabilities, leading to the collapse of gas clouds and the formation
of new stars. The tidal interactions are more effective in inducing
star formation during this period due to the lack of stability in the
galaxies \citep{tissera02}. As the blue galaxies evolve, their star
formation activity can deplete the available gas, leading to a decline
in stellar birth rates. The transition from blue to red occurs as
galaxies age and star formation becomes less vigorous.

Besides natural aging, the transformation of blue galaxies into red
galaxies involve a complex interplay of various astrophysical
processes. The fact that the red and the blue galaxies preferentially
reside in high-density and low-density environments respectively
\citep{hogg04, baldry04b, balogh, blan05, park05, pandey05, zehavi05,
  pandey06, bamford09, cooper10, pandey20, nandi23}, suggests that the
environment plays a decisive role in such transformation.

The galaxies accumulate their gas from the cosmic streams
\citep{dekel06} and the circumgalactic medium \citep{maller04}. The
galaxies can also reach a quiescent stage if some physical processes
prevent gas accumulation or suppress the star formation
efficiency. The expulsion or removal of gas from the galaxy can be
another efficient route for quenching star formation. The ram pressure
stripping \citep{gunn72} in the high-density environment and the gas
loss due to feedback from supernovae, active galactic nuclei, or
shock-driven winds \citep{cox04, murray05, springel05} can drastically
suppress the star formation in galaxies. Galactic interactions and
mergers provide other possible routes for transformation. A seminal
study of optical colours in morphologically disturbed galaxies by
\cite{larson78} provided the first observational evidence of SFR
enhancement in interacting galaxies. Subsequently, numerous
corroborative studies \citep{ barton00, lambas03, alonso04, nikolic04,
  woods06, woods07, barton07, ellison08, ellison10, woods10, patton11,
  barrera15, thorp22, shah22, das22} employing spectroscopic redshift
surveys have provided abundant supporting evidence, consistently
reaffirming this phenomenon. The SFR enhancement in interacting
galaxies is known to be influenced by several factors such as
proximity, luminosity or mass disparity, and the specific types of
galaxies involved in the interaction. These observational findings are
consistent with various studies on galaxy interaction using
simulations. A pioneering study using the simulation of galaxy
interactions by \cite{toomre72} first showed that spiral and irregular
galaxies could transform into ellipticals and S0 galaxies. Subsequent
research utilizing advanced simulations \citep{barnes96, mihos96,
  tissera02, cox06, montuori10, lotz11, torrey12, hopkins13, renaud14,
  renaud15, moreno15, moreno21, renaud22} has further elucidated that
tidal torques induced during galactic encounters can instigate
episodes of intense star formation, known as starbursts, within
interacting galaxies. The effectiveness of these tidal-induced
starbursts depends on multiple factors such as the morphology, the
depth of the gravitational potential well, the amount of available
gas, orbital characteristics, and the internal dynamics of the
galaxies \citep{barnes96, tissera00, perez06}.

Thus, interactions between galaxies can trigger starbursts and alter
the structure of the merging galaxies. The aftermath of such mergers
can give rise to red and blue galaxies, depending on the specific
conditions and the available gas reservoirs. For instance, the
interactions between gas-rich galaxies can enhance the star formation
activity, turning them bluer. On the other hand, the mechanisms such
as strangulation \citep{gunn72, balogh00}, harassment \citep{moore96,
  moore98}, starvation \citep{larson80, somerville99, kawata08},
merger \citep{hopkins08} and quenching of satellites \citep{geha12}
can cease star formation in galaxies and make them redder. Some
secular processes, such as morphological quenching \citep{martig09}
and bar quenching \citep{masters10}, can also quench star formation
and alter the colour of a galaxy over long timescales.

Observations indicate a peak in the star formation rates of galaxies
somewhere between the redshift of $2-3$ \citep{tran10, forster20,
  gupta20}. This epoch is usually referred to as the cosmic noon. A
sharp decline in the cosmic star formation rate between $z=1$ to $z=0$
\citep{madau96} suggests significant evolution of galaxy properties in
the recent past. Understanding the physical processes responsible for
the global decline in the star formation activity in galaxies is
crucial to our understanding of the galaxy evolution.

The Evolution and Assembly of Galaxies and their Environment (EAGLE)
simulations \citep{schaye15} is a suite of cosmological hydrodynamical
simulations designed to study the formation and evolution of galaxies
within a cosmological context. The simulations achieve high spatial
and mass resolution, allowing for the detailed study of small-scale
processes within galaxies, such as star formation, feedback, and
interactions. The simulation provides the observed properties of
galaxies across a wide range of cosmic epochs. It reproduces the
observed diversity of galaxies in the real universe and helps us to
understand the underlying physical processes that drive their
evolution \citep{furlong15, crain15, trayford16}.

Several earlier works study quenching in galaxies using different
hydrodynamical simulations \citep{furlong15, trayford16, wright19,
  donnari21, walters22}. In this work, we plan to use the data from
the EAGLE simulation to understand the roles of environment and
interactions in the transformation of galaxy colour since the redshift
of $\sim 3$. The snapshots from different redshifts between $z=0-3$
will be used to study the evolution. We will classify the galaxies
into red and blue populations based on their colour and stellar mass
using a recently proposed classification scheme \citep{pandey23}.  The
environment of the galaxies can be characterised by their local
density. One can also use the two-point correlation function to
measure the clustering strength. The comparisons of the environment
and the clustering strength of the red and blue galaxies at different
redshifts would allow us to understand their roles in the galaxy
evolution. The galaxy interactions may also have a role in leading the
migration of the galaxies from the blue cloud to the red sequence. We
can study the roles of interactions in galaxy evolution by comparing
the properties of the paired galaxies across different redshifts. At
each redshift, we plan to identify the interacting galaxies and
prepare the respective samples of the isolated galaxies by matching
their stellar mass and environment. We want to compare the star
formation rates of the interacting galaxies and their isolated
counterparts throughout the entire redshift range. The fraction of the
red and blue galaxies in the interacting population across different
redshifts can also reveal the roles of interactions in the
transformation of galaxy colour.

The paper is organised as follows: the data is described in Section 2,
the method of analysis is outlined in Section 3, the results are
discussed in Section 4, and the conclusions are presented in Section
5.

\section{Data}

We use the data from the EAGLE simulations \citep{schaye15,mcalpine16,
  crain15} for the present analysis. EAGLE incorporates sophisticated
hydrodynamical models that simulate the behaviour of gas, stars, and
dark matter in a cosmological context. It follows the evolution of
both baryonic and non-baryonic components of matter from the redshift
of $z=127$ to the present. The simulation generates the populations of
galaxies at different cosmic epochs that exhibit a diverse range of
properties, including sizes, masses, colours, and star formation
rates.  This information is available ranging from a lookback time of
13.62 Gyr ($z=20$) to present ($z=0$). EAGLE adopts a flat
$\Lambda$CDM cosmology with parameters $\Omega_{\Lambda}=0.693,
\Omega_m=0.307, \Omega_b=0.04825, H_0=67.77$ km/s/Mpc
\citep{planck14}.

We download various physical properties of galaxies within a cubic
volume that extends to 100 comoving Mpc on a side. The spatial
distributions of the galaxies and their physical properties are
downloaded for the following redshifts $z = 0, 0.5, 1, 1.49, 2.01,
2.48, 3.02$, which respectively corresponds to lookback times $0,
5.19, 7.93, 9.49, 10.53, 11.16, 11.66$ (in Gyrs). We extract the
information of the location of the minimum gravitational potential of
galaxies from $Ref-L0100N1504\_Subhalo$ table. We select only the
genuine simulated galaxies by using the condition $Spurious=0$. This
condition discards all the unusual objects with anomalous stellar
mass, metallicity or black hole mass. We use table
$Ref-L0100N1504\_Aperture$ to download the star formation rate,
stellar mass and cold gas mass of the galaxies. These properties of
the simulated galaxies are estimated within a spherical 3D aperture
with a radius of 30 kpc centered around the minimum of their
gravitational potential. The properties estimated within this aperture
size are found to be well suited for comparison with observations
\cite{schaye15}. Further, we consider only galaxies with stellar mass
> 0. We combine the two tables using $GalaxyId$ to obtain the above
mentioned information. This criteria gives us $325358$ galaxies with
the required information for redshift $z=0$. Further, we use table
$Ref-L0100N1504\_Magnitude$ to extract the information of rest frame
broadband magnitudes of galaxies estimated in $u$ and $r$ band filters
\cite{doi10}. Here $u$ and $r$ band respectively denotes the
Ultraviolet and Red filters used in the Sloan Digital Sky Survey
(SDSS). The magnitudes of the galaxies are also computed within 30 kpc
spherical aperture at the location of minimum gravitational potential
\cite{trayford15}. We finally combine three tables using $GalaxyID$ to
retrieve this information. It is important to note that we obtain the
stellar mass estimates from the $Ref-L0100N1504\_Aperture$ table,
which includes information solely for galaxies with stellar masses
$log(M_{stellar}/M_{sun}) > 8.3$. The baryonic particle mass for this
simulation is $1.81 \times 10^6 M_{sun}$. This criterion also ensures
that each galaxy included in our analysis is sufficiently resolved,
with at least approximately $110$ particles. We find that for redshift
$z=0$ there are $29754$ galaxies with the required information. We
observe a notable reduction in the galaxy count upon merging different
tables, primarily due to the minimum threshold in stellar mass
estimates in $Ref-L0100N1504\_Aperture$ table and the inclusion
criteria limited to galaxies with available rest-frame broadband
magnitudes in the $u$ and $r$ band filters listed in
$Ref-L0100N1504\_Magnitude$ table. For the redshifts $z=0.5, 1, 1.49,
2.01, 2.48, 3.02$, we have $31740, 31357, 28745, 24074, 19288,$ and
$14533$ galaxies respectively.


\section{Method of analysis}
Our primary goal in this work is to classify galaxies in the Eagle
simulation into red and blue populations based on stellar mass and
redshift, and then study the roles of interactions and environments in
their evolution.

\begin{figure*}[htbp!]
\centering
\includegraphics[width=15cm]{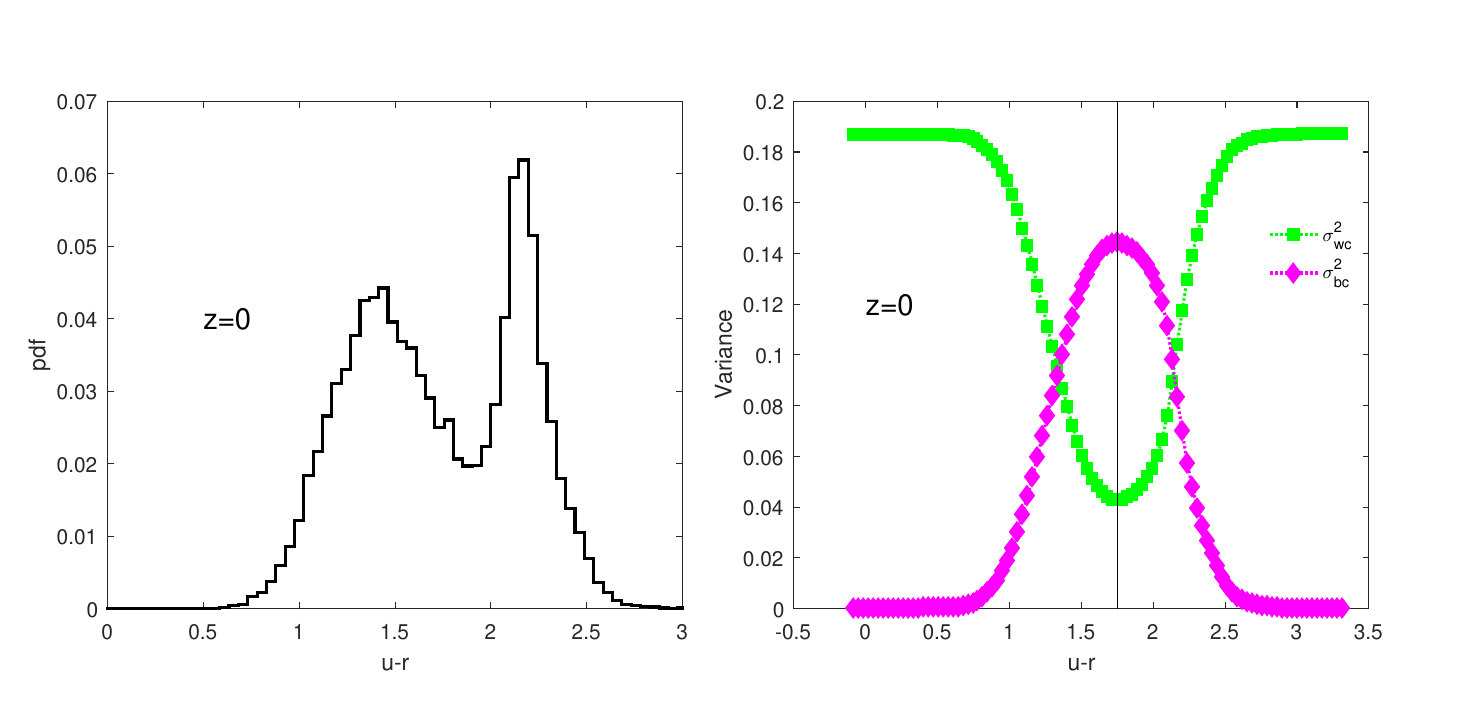}

\includegraphics[width=15cm]{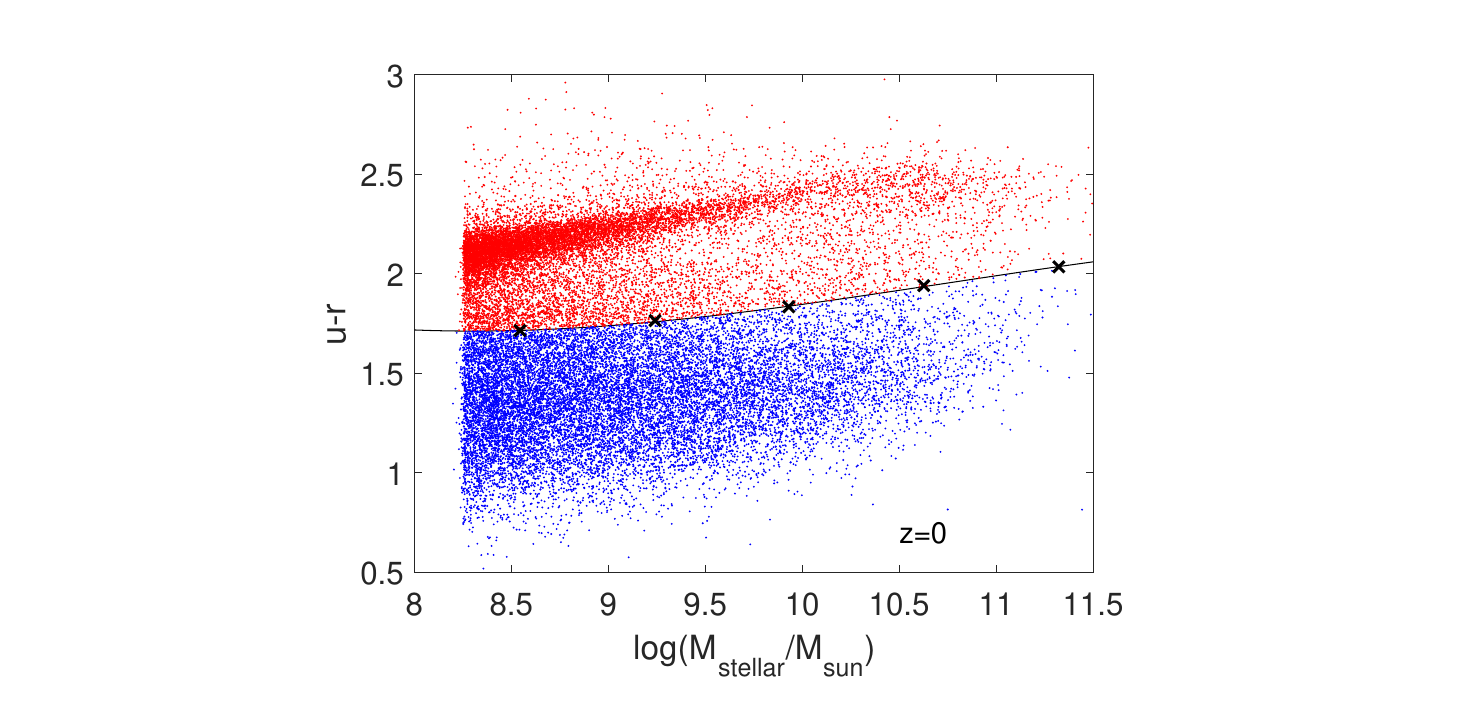}
\caption{The top left panel shows the ($u-r$) colour distribution of
  galaxies. The top right panel shows the within-class variance
  $\sigma^2_{wc}$ and between-class variance $\sigma^2_{bc}$ of the
  entire sample as function of $u-r$ color. The vertical black line
  shows the ($u-r$) colour threshold for separating the red and blue
  galaxies. The bottom panel shows the red (red dots) and blue
  galaxies (blue dots) in the colour-stellar mass plane. The black
  crosses represent the threshold colour obtained in different stellar
  mass bins. The solid black line is obtained by fitting a cubic
  polynomial to the threshold values. The results shown here are for
  redshift $z=0$.}
\label{Fig1}
\end{figure*}

\subsection{Classifying the red and blue galaxies}

We classify the galaxies as red and blue using a scheme based on the
Otsu's method \cite{pandey23}. Otsu's technique \cite{otsu79} was
initially proposed to separate the foreground pixels from the
background pixels in a grayscale image based on their intensities. The
pixels with intensities greater than a threshold value are marked as
foreground. The remaining pixels with intensities less than or equal
to the threshold are labelled as background. This method is most
suitable for a bimodal distribution of pixel intensities. It converts
the input grayscale image into a binary image with foreground and
background pixels.

The $(u-r)$ colour distribution of galaxies shows a bimodal nature
with two distinct peaks \cite{strateva01,balogh04,baldry04}. The
bimodality in colour is also evident for the galaxies in the EAGLE
simulation (\autoref{Fig1}). The classification scheme used in this
work is ideal for any such distribution. We apply this technique to
the $(u-r)$ colour distribution of the galaxies from the EAGLE
simulation at $z=0$. We first calculate the histogram of $(u-r)$
colour of galaxies with $M$ number of bins. \cite{pandey23} show that
the classification scheme is independent of the choice of number of
bins. We choose $M=100$ for the present analysis. We then normalize
the histogram by dividing the number of galaxies in each bin by the total
number of galaxies N. The probability of galaxies $p_i$ associated
with the $i^{th}$ bin is then given by $p_i=\frac{n_i}{N}$. Here $n_i$
represents the number of galaxies in the $i^{th}$ bin. The resulting
probability distribution of galaxies is then used to calculate the
probabilities of class occurrence for the red sequence and the blue
cloud. If $k^{th}$ bin in the histogram corresponds to the threshold,
then bins $[1,....,k]$ belong to the blue cloud and $[k+1,....,M]$
belong to the red sequence. If $B$ and $R$ respectively denote blue
cloud and red sequence, the probability of class occurrence for the two
populations are respectively given by,

\begin{equation}
P_B=\sum^k_{i=1}p_i
\end{equation}

\begin{equation}
P_R=\sum^M_{i=k+1}p_i
\end{equation}

The class means for each threshold for the two class $B$ and $R$ are
given by,

\begin{equation}
\mu_B=\frac{\sum^k_{i=1}c_ip_i}{P_B}
\end{equation}

\begin{equation}
\mu_R=\frac{\sum^M_{i=k+1}c_ip_i}{P_R}
\end{equation}

Here $c_i$ represents the mean $(u-r)$ colour corresponding to the $i^{th}$ bin. The class variances for $B$ and $R$ are then respectively given by,

\begin{equation}
\sigma^2_B=\frac{\sum^k_{i=1}(c_i-\mu_B)^2p_i}{P_B}
\end{equation}

\begin{equation}
\sigma^2_R=\frac{\sum^M_{i=k+1}(c_i-\mu_R)^2p_i}{P_R}
\end{equation}

The within-class variance $\sigma^2_{wc}$ and between-class variance
$\sigma^2_{bc}$ can be respectively written as,

\begin{equation}
\sigma^2_{wc}=P_B\sigma^2_B+P_R\sigma^2_R
\label{eq:wc}
\end{equation}

\begin{equation}
\sigma^2_{bc}=P_BP_R(\mu_B-\mu_R)^2
\label{eq:bc}
\end{equation}

Otsu's method iteratively searches for the threshold that minimizes
the within-class variance or maximizes the between-class variance. We
show the within-class variance (\autoref{eq:wc}) and between-class
variance (\autoref{eq:bc}) of the entire sample as a function of
$(u-r)$ colour threshold in the right panel of \autoref{Fig1}. It is
clearly seen that $\sigma^2_{wc}$ is minimum and $\sigma^2_{bc}$ is
maximum at the same $(u-r)$ colour threshold. We consider this optimum
$(u-r)$ colour threshold for separating the galaxies into the ``blue
cloud'' and the ``red sequence''. We classify the galaxies with a
colour greater than the threshold as red. Similarly, galaxies whose
colour is less than the threshold are classified as blue. It is well
known that the galaxy colour depends on the stellar mass. The stellar
mass of the galaxies varies over a wide range, and one can not apply a
single colour threshold to divide the galaxies into the red and blue
populations. We segregate the galaxies according to their stellar mass
into different mass bins. We then apply the same technique to the
galaxies in different mass bins. It provides a separate $(u-r)$ colour
threshold for each mass bin. The colour threshold obtained as a
function of the stellar mass is shown in the bottom panel of
\autoref{Fig1}. We find that the colour threshold increases with
increasing stellar mass. We fit the threshold values as a function of
the stellar mass with a smooth cubic polynomial as shown in
\autoref{Fig1}. The classified galaxies and the line separating them
are shown in the colour-stellar mass plane in \autoref{Fig1}. The
results shown in \autoref{Fig1} correspond to the galaxies in the
simulation at $z=0$. We apply the same technique to the simulation
outputs at different redshifts and classify the red and blue galaxies
analogously.

\subsection{Local environment}

We calculate the local density $(\eta_k)$ at the location of each
galaxies using the $k^{th}$ nearest neighbour method
\cite{casertano85}. The local density $\eta_k$ given by,

\begin{equation}
\eta_k=\frac{k-1}{\frac{4}{3}\pi r^3_k}
\end{equation}

Here $r_k$ is the distance from the galaxy to its $k^{th}$ nearest
neighbour. We consider the distance to the $5^{th}$ nearest neighbour
by taking $k=5$ in this analysis.


\subsection{Clustering strength}

We quantify the clustering strength of the galaxies in our sample
using the two point correlation function $\xi(r)$. The two point
correlation function $\xi(r)$ traces the amplitude of clustering
strength as function of length scale $r$. $\xi(r)$ is a measure of the
excess probability of finding two galaxies at a given separation $r$
compared to an unclustered random Poisson distribution. We use the
Peebles \& Hauser estimator \cite{peebles74} for measuring $\xi(r)$.
The $\xi(r)$ is given by,

\begin{equation}
\xi(r)=\frac{DD(r)}{RR(r)}-1
\end{equation}

Here $DD(r)$ and $RR(r)$ respectively represents the number of galaxy
pairs in data and random Poisson distribution for separation $r$.


It is well known that $\xi(r)$ for galaxies exhibit a power law behaviour \cite{peebles75} ,

\begin{equation}
\xi(r)=\left(\frac{r}{r_o}\right)^{-\gamma}
\end{equation} 

where $r_0$ represents the correlation length.

\subsection{Identifying galaxy pairs and building control samples of isolated galaxies}

We identify the galaxy pairs in the EAGLE simulation following the
strategy described in Das et al. (2023) \cite{das23}. We consider a
galaxy and its first nearest neighbour as a pair if their separation
in real space $r \leq 200 $ kpc. Here, $r$ refers to the
three-dimensional separation between the locations of their minimum
gravitational potential. We identify the paired galaxies at redshift
$z = 0, 0.5, 1, 1.49, 2.01, 2.48, 3.02$. We also prepare control
samples of isolated galaxies for the identified pairs at redshift $z =
0, 1, 2.01, 3.02$. We identify the isolated galaxies in the simulation
volume as those that do not have a neighbour within a radius of 1 Mpc
around them. We prepare a control sample of isolated galaxies by
matching their stellar mass and local density with the paired
galaxies. The isolated galaxies are matched to the paired galaxies
within $\Delta_{mass}$ =0.08 dex and $\Delta_{\eta_5}= 0.006$
$Mpc^{-3}$. We compare the distributions of stellar mass and the local
density of the pair and control-matched sample by applying
Kolmogorov-Smirnov (KS) test. The values of $\Delta_{mass},$
$\Delta_{\eta_5}$ are chosen in such a way that the null hypothesis
for the distributions of stellar mass, local density of the pair and
the control sample can be rejected at $\leq 40\%$ confidence level. It
ensures that the distributions of stellar mass and local density of
the paired and isolated galaxies are highly likely to be drawn from
the same parent population.

The maximum stellar mass ratio of galaxy pairs in our sample can be as
extreme as $1:1000$. Any interactions would have a negligible effect
on the more massive members of the interacting pairs with such extreme
mass ratio. For the present analysis, we only consider the interacting
galaxy pairs with stellar mass ratio $1 \leq \frac{M_1}{M_2} \leq
10$. This ensures that the galaxy pairs in our sample are more likely
to undergo noticeable changes during or after close
interactions. Further, it becomes difficult to accurately disentangle
galaxy pairs in simulations \citep{rod15, mcalpine20, patton20}. We
impose a minimum pair separation ($r \geq 30$ kpc) to mitigate this
issue.

The galaxy pairs are identified based on the three-dimensional
physical separation between the galaxies in real space. Some of the
identified pairs may belong to the galaxy groups or clusters. To avoid
these pairs, we also impose an additional condition that the pair
members must have their second nearest neighbour at a distance $>500$
kpc. We repeat our pair search and identify the set of galaxy pairs
from the different snapshots between $z=0-3$. At each redshift, the
control samples of these paired galaxies are also prepared from the
samples of the isolated galaxies by matching their stellar mass and
environment.


\begin{figure*}[htbp!]
\centering
\includegraphics[width=12cm]{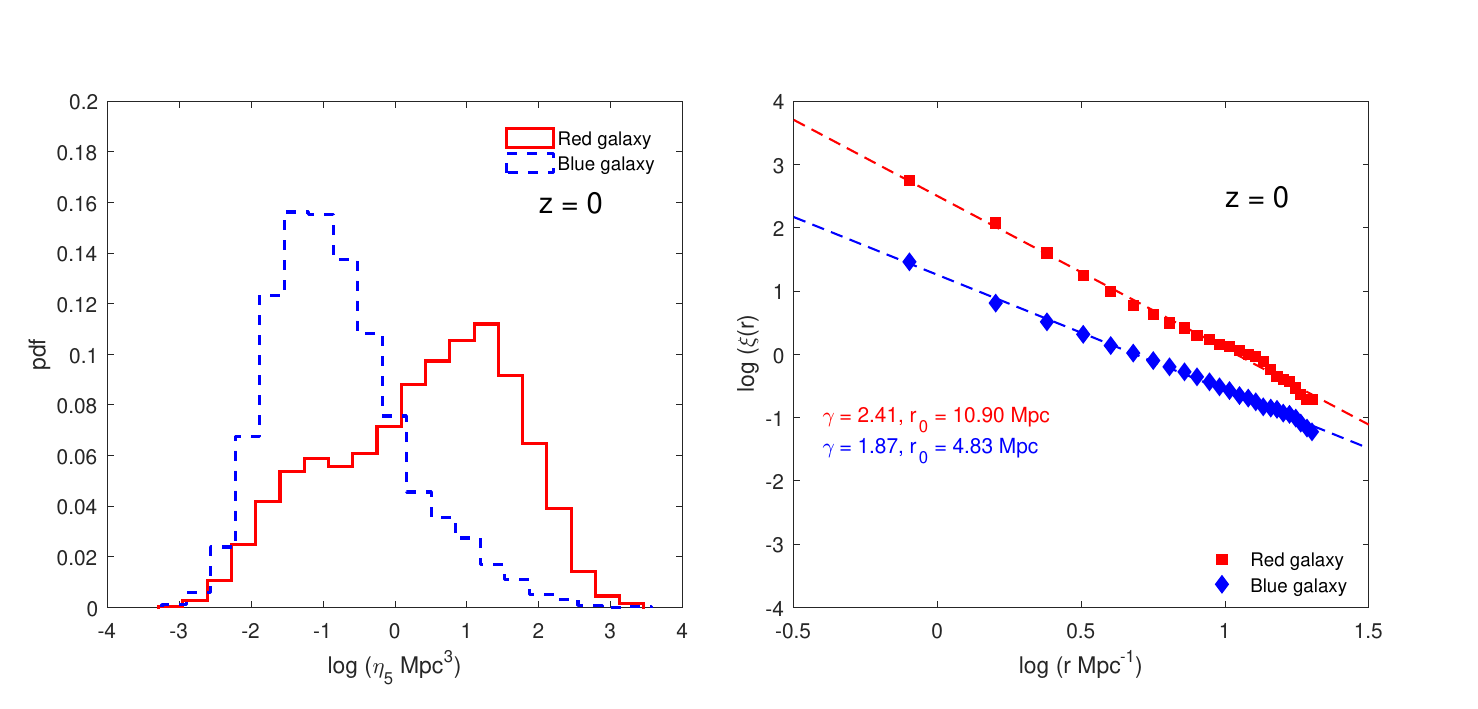}
\includegraphics[width=12cm]{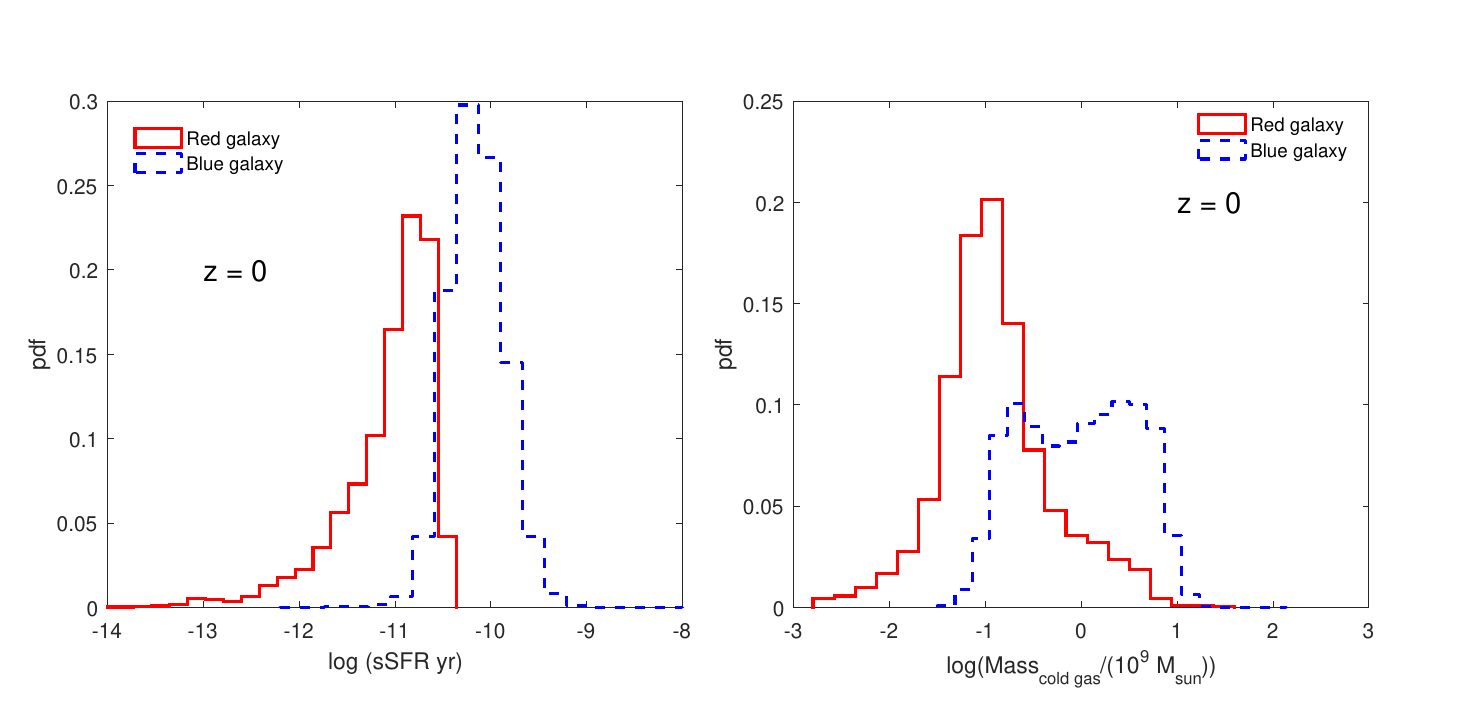}
\includegraphics[width=12cm]{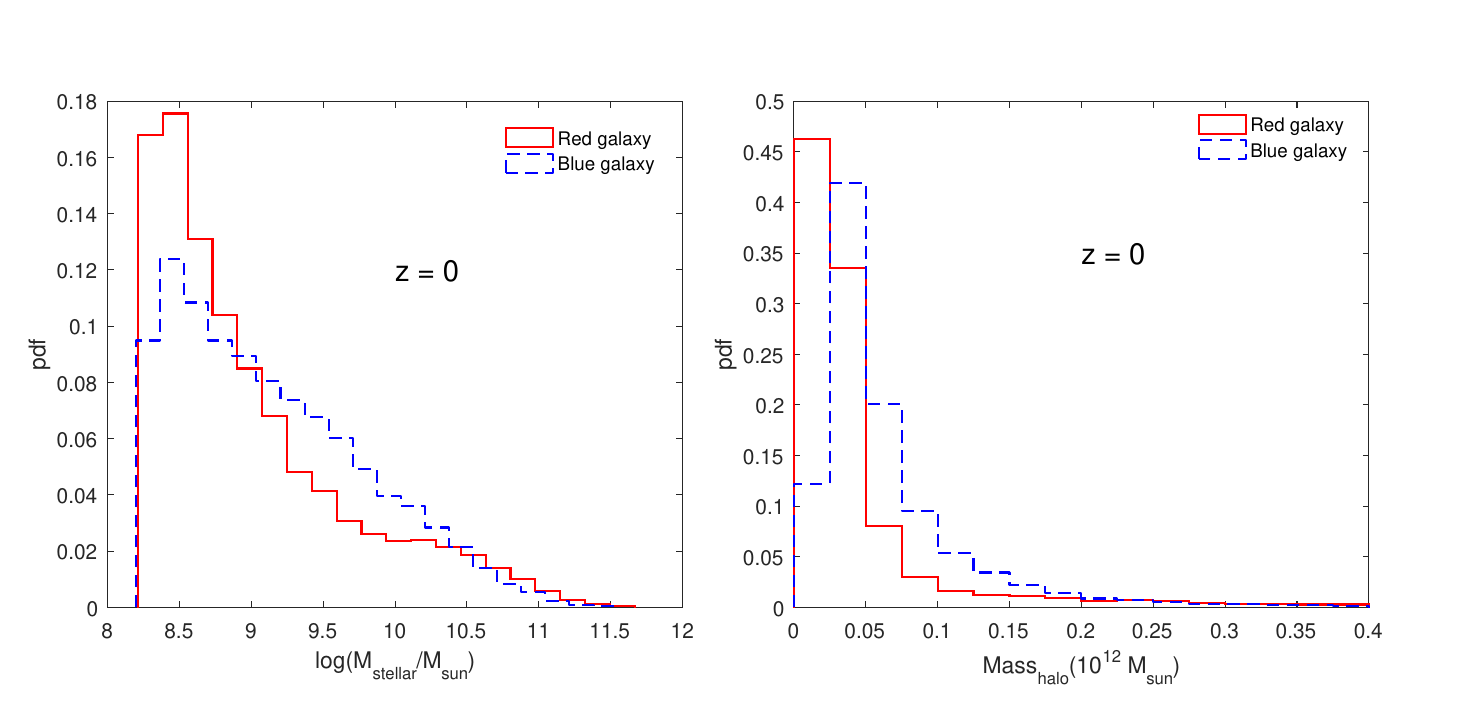}
\caption{The top left panel compares the PDFs of the local density of
  the red and blue galaxies at redshift $z=0$. The two-point
  correlation functions of the red and blue galaxies at the same
  redshift are compared in the top right panel. The best fit lines to
  the correlation functions and the fitted parameters are shown
  together in the same panel. The middle left, middle right, bottom
  left and the bottom right panels respectively compare the
  distributions (PDFs) of the sSFR, cold gas mass, stellar mass and
  halo mass of the red and blue galaxies at $z=0$.}
\label{Fig2}
\end{figure*}

\begin{figure*}[htbp!]
\centering
\includegraphics[width=14cm]{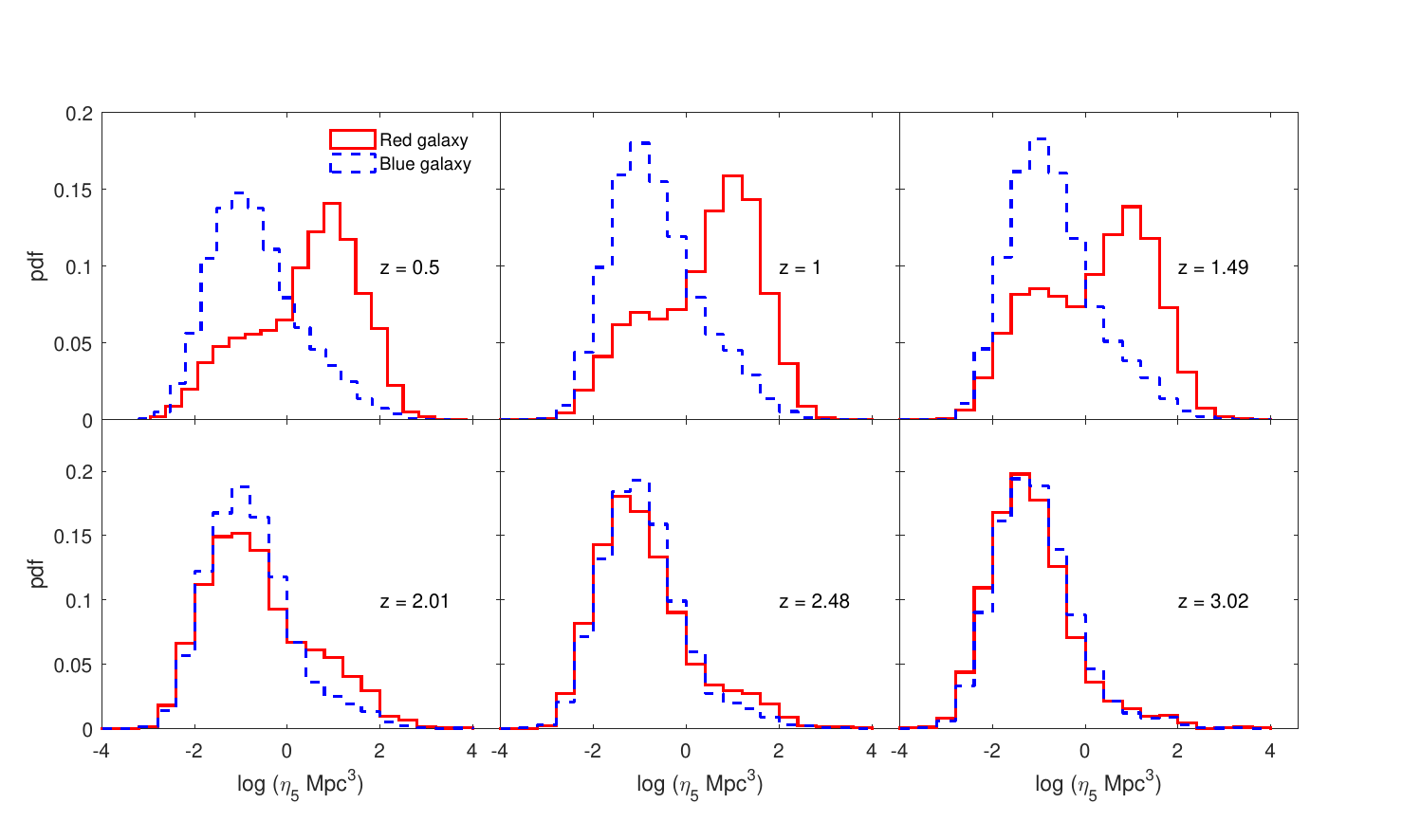}
\caption{The different panels compares the PDFs of the local density
  of the red and blue galaxies at different redshifts between
  $z=0.5-3.02$.}
\label{Fig3}
\end{figure*}

\begin{figure*}[htbp!]
\centering
\includegraphics[width=14cm]{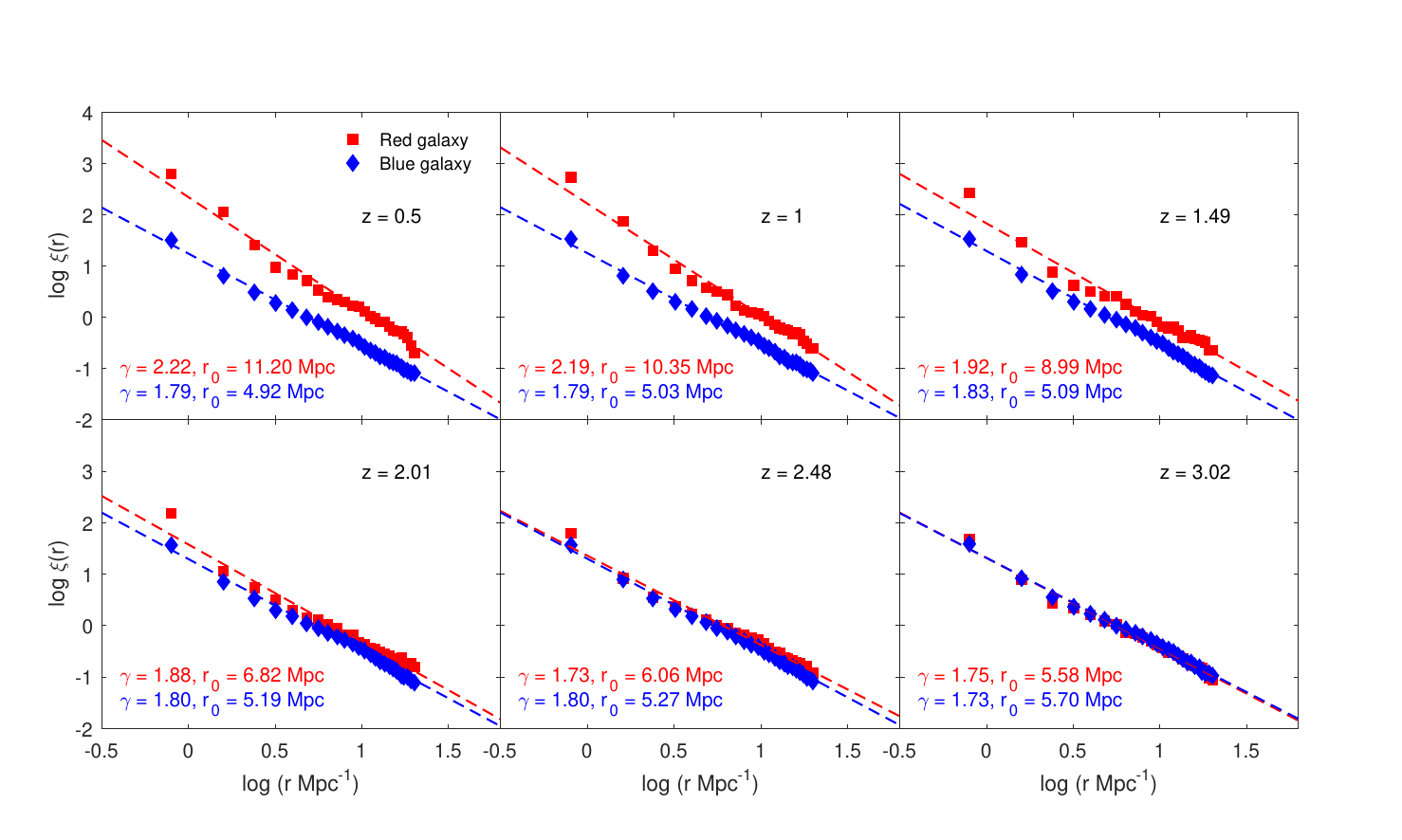}
\caption{The different panels compare the two-point correlation
  function of the red and blue galaxies at different redshifts between
  $z=0.5-3.02$. The best fit lines and the associated parameters are
  shown together in each panel.}
\label{Fig4}
\end{figure*}

\begin{figure*}[htbp!]
\centering
\includegraphics[width=14cm]{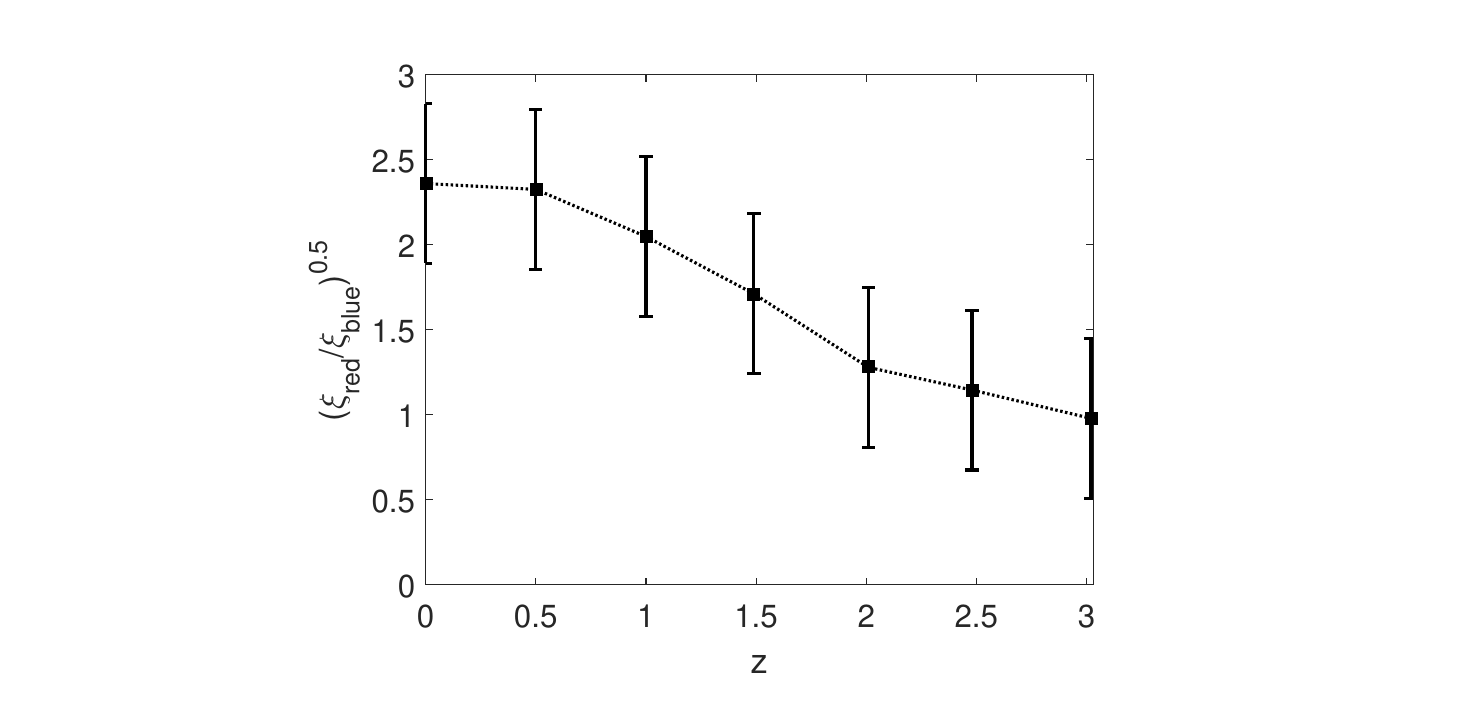}
\caption{This shows the relative bias of the red and blue galaxies as
  a function of redshift between $z=3.02-0$. The 1$\sigma$ error bars
  shown at each data point are obtained from 10 jackknife samples
  drawn from the original datasets.}
\label{bias}
\end{figure*}

\begin{figure*}[htbp!]
\centering
\includegraphics[width=14cm]{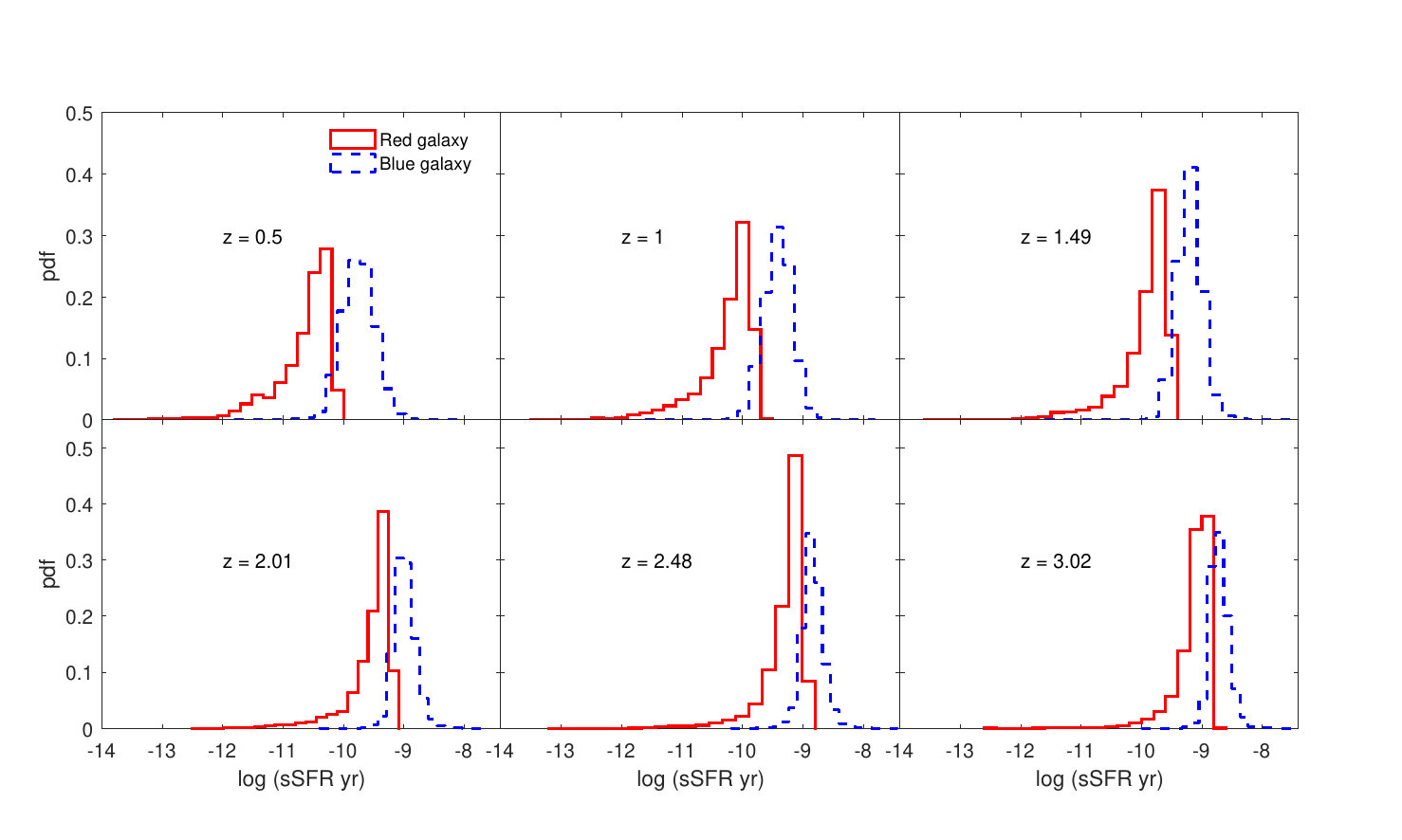}
\caption{Same as \autoref{Fig3}, but for sSFR.}
\label{Fig5}
\end{figure*}

\begin{figure*}[htbp!]
\centering
\includegraphics[width=14cm]{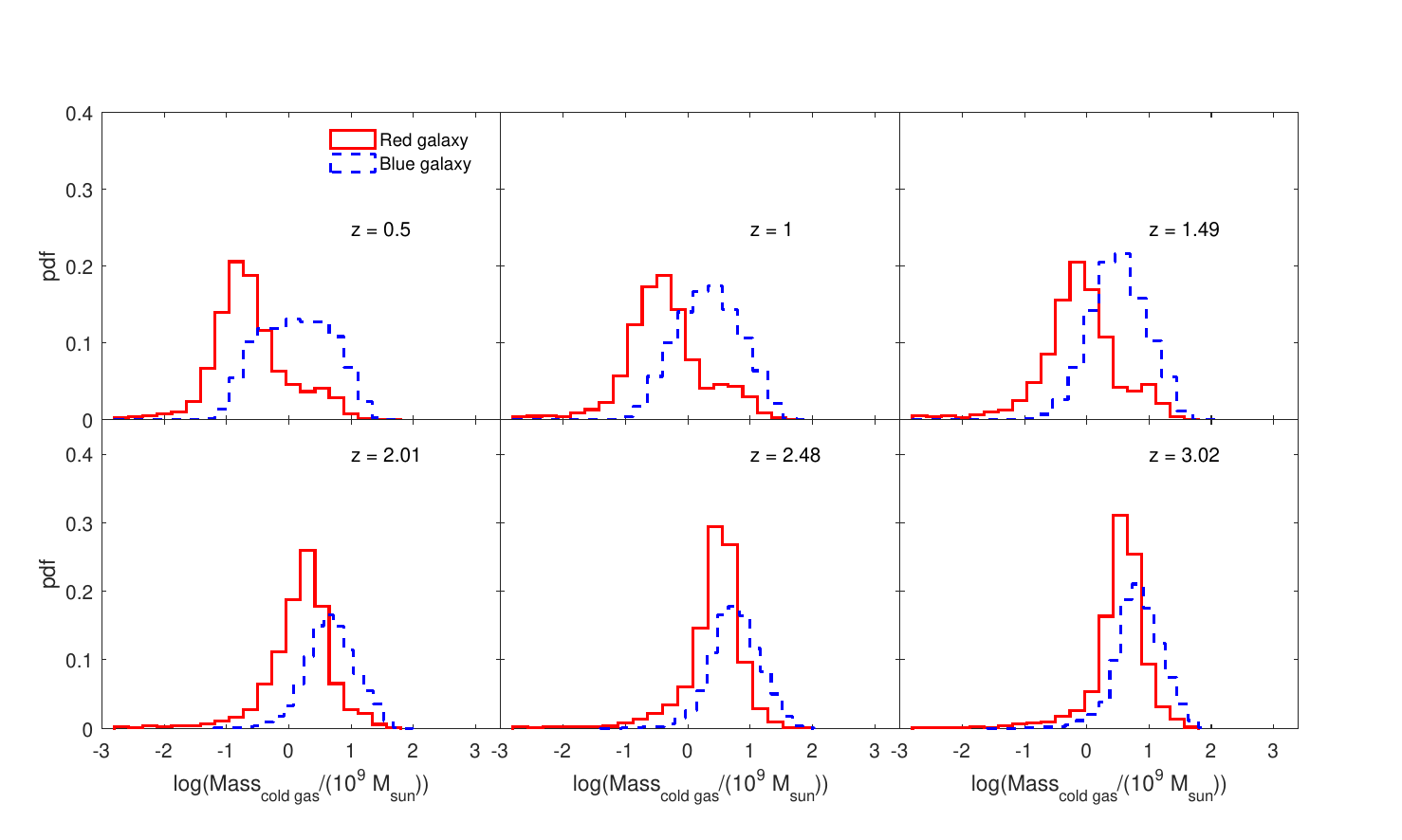}
\caption{Same as \autoref{Fig3}, but for cold gas mass.}
\label{Fig6}
\end{figure*}

\section{Results}

We first compare the local density, clustering strength, sSFR, cold
gas mass, stellar mass and the dark matter halo mass of the red and
blue galaxies at the present epoch. The results are shown in different
panels of \autoref{Fig2}.

The top left panel of \autoref{Fig2} compares the PDF of the local
density of the red and blue galaxies in the present universe. It
clearly shows that the blue galaxies primarily reside in the
low-density environments, whereas the red galaxies dominate the
high-density environments. A comparison of the two-point correlation
function of the red and blue galaxies in the top right panel of
\autoref{Fig2} shows that the red galaxies are strongly clustered
compared to the blue galaxies. On average, the clustering strength of
the red galaxies is $\sim 6$ times stronger compared to the blue
galaxies within $r<20$ Mpc. Also, the correlation length ($r_{0}$) of
the red galaxies is more than twice that of the blue galaxies. The
galaxies residing in the high-density regions are expected to be more
strongly clustered. The fact that the red galaxies are preferentially
located in the high-density environment is consistent with their
stronger clustering strength. The middle left and right panels of
\autoref{Fig2} compare the sSFR and cold gas mass distributions of the
red and blue galaxies at $z=0$. These show that, in general, the
galaxies in the present universe have a low cold gas mass
reservoir. Notably, the red galaxies have nearly depleted their entire
cold gas mass reservoir, resulting in negligible or no star
formation. The middle left panel shows that the blue galaxies
generally exhibit higher SSFR compared to red galaxies due to their
ongoing or recent star formation activity, which is typically
sustained by the availability of gas reservoirs for forming new
stars. In contrast, red galaxies, which are often older and more
evolved, have lower specific star formation rates because they have
depleted much of their gas reservoirs and thus form stars at a slower
rate. It would also be interesting to compare the stellar and halo
mass distributions of the red and blue galaxies at present. The bottom
left panel of \autoref{Fig2} compares the PDF of the stellar mass
distributions of the red and blue galaxies at $z=0$. Interestingly,
the red population dominates the blue population both at the low and
the high end of the stellar mass distributions. The red galaxies at
the low stellar mass end could be the quenched satellite galaxies,
whereas those at the high stellar mass end could be the massive
central galaxies that have quenched their star formation through mass
quenching or AGN feedback. We find an analogous behaviour in the halo
mass distribution of the red galaxies as shown in the bottom right
panel of \autoref{Fig2}. A large fraction of the galaxies in the red
population are hosted in the smaller mass halos. These smaller mass
halos may correspond to the low-mass satellite galaxies.

We now study the evolution of the properties of the red and blue
galaxies as a function of redshift. We compare the PDF of the local
density of the red and blue galaxies at different redshifts in the
different panels of \autoref{Fig3}. The results show that the local
environment of the red and blue galaxies have smaller differences up
to $z=2.01$. The differences in their local environment increase
progressively with decreasing redshift. The number of blue galaxies
dominates the low-density environment, whereas the red galaxies
progressively dominate the high-density environments. A comparison of
the two-point correlation function of the red and blue galaxies across
different redshifts in \autoref{Fig4} confirm the same trend. It shows
that the red and blue galaxies have similar clustering strength up to
$z=2.48$. The red galaxies cluster more strongly than the blue
galaxies after $z=2.01$. The differences in the values of $\gamma$ and
$r_{0}$ for the red and blue galaxies increase with time. We show the
relative bias between the red and blue galaxies as a function of
redshift in \autoref{bias}. \autoref{bias} shows that the red and blue
galaxies have the same bias at $z \sim 3$. The relative bias
proliferates after $z=2$, reaching a value of $\sim 2.5$ at
$z=0$. These results indicate that the environment may not entirely
decide the transformation of galaxy colour before $z=2$. The
environmental factors dominate at smaller redshifts ($z<2$), leading
to significant evolution in galaxy properties.

We compare the sSFR distributions of the red and blue galaxies at
different redshifts in \autoref{Fig5}. The distributions of the cold
gas mass in these two populations are compared at different redshifts
in \autoref{Fig6}. Both \autoref{Fig5} and \autoref{Fig6} confirm a
strong correlation between the sSFR and the cold gas mass in both the
red and blue galaxies. Interestingly, the sSFR and cold gas mass
distributions of the red and blue galaxies are significantly different
between $2 \leq z \leq 3$ even though their local environment and the
clustering strength have smaller differences during this period
(\autoref{Fig3} and \autoref{Fig4}). It implies that the dichotomy in
the star formation histories of galaxies does not originate only from
the differences in their local density. The size of the cold gas mass
reservoir in galaxies may be related to the gas accretion efficiency
of their host dark matter halos. The greater abundance of the cold gas
mass in some galaxies may lead to higher sSFR. At lower redshifts
$z<2$, the environmental factors like starbursts during galaxy
interactions \citep{barton00, lambas03, alonso04, nikolic04, woods06}
boost the conversion efficiency of gas to stars, eventually depleting
their cold gas reservoir. Several other environment-driven physical
mechanisms such as ram pressure stripping \citep{gunn72}, harassment
\citep{moore96}, starvation \citep{larson80} and feedback from
supernovae and AGN \citep{cox04, murray05} can prevent the gas
replenishment or expel the gas from the galaxy, eventually quenching
the star formation.

\begin{figure*}[htbp!]
\centering
\includegraphics[width=14cm]{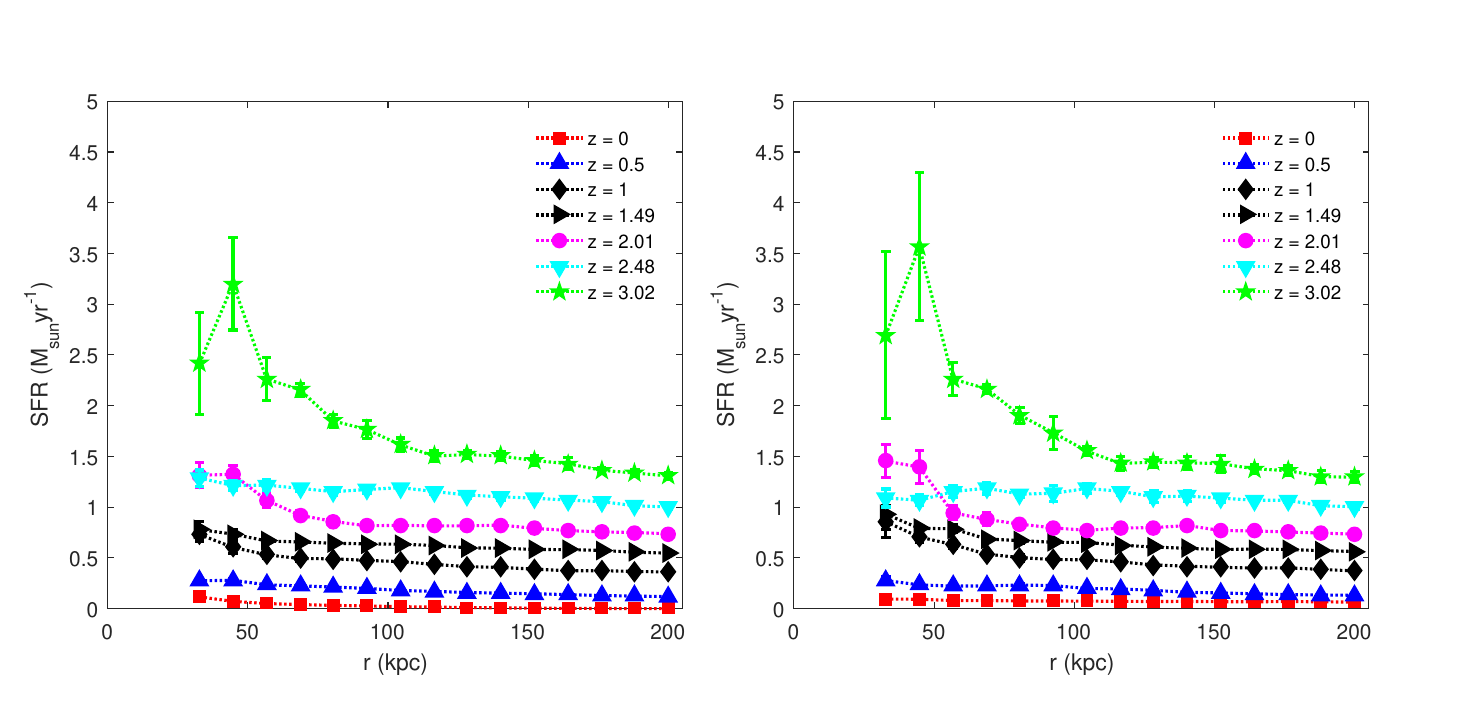}
\includegraphics[width=14cm]{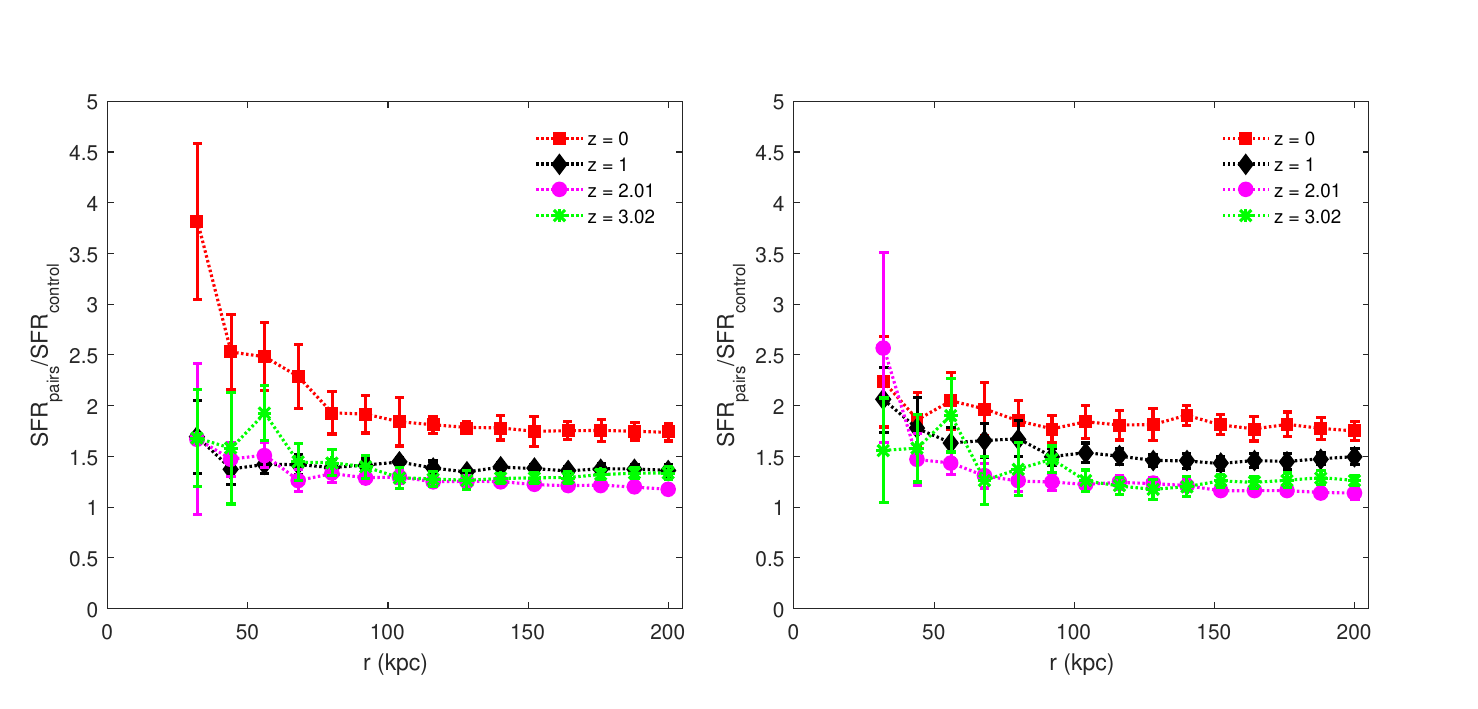}
\caption{The top left panel shows the cumulative median SFR of all
  paired galaxies as a function of the pair separation at different
  redshifts between $z=3.02-0$. The ratio of the cumulative median SFR
  of the control-matched paired galaxies and isolated galaxies is
  shown as a function of the pair separation at different redshifts in
  the bottom left panel. Only the pairs having $1 \leq \frac{M_1}{M_2}
  \leq 10$ and separation $r \geq 30$ kpc are considered in this
  analysis. The top right and bottom right panels show the same but
  for the paired galaxies, which have their second nearest neighbour
  at a distance $>500$ kpc. The 1$\sigma$ errors shown at each data
  point are obtained from $10$ jackknife samples drawn from the
  original data.}
\label{interact1}
\end{figure*}

The galaxy interactions play an important role in the evolution of the
galaxy properties. The interactions may significantly impact the
colour of a galaxy and its evolution. We compare the cumulative median
SFR of the paired galaxies as a function of the pair separation at
different redshifts in the top left panel of \autoref{interact1}. The
paired galaxies have a higher median SFR at a given pair separation in
the past. This is related to the increasing gas fractions in all
galaxies at higher redshift. Such trends have been reported in earlier
studies \citep{hopkins10, forster20}. At each redshift, we observe a
somewhat higher median SFR in interacting galaxies at smaller pair
separation. This rise in SFR at smaller pair separations is most
likely linked to the interaction between the member galaxies in paired
systems. An anomalous behaviour is observed at $z= 2.01$ where the
paired galaxies with separation $<50$ kpc exhibit a comparable or
higher median SFR than those with the same separation at $z=2.48$. It
indicates a surge in close galaxy interactions around $z\sim 2$, which
induced tidally triggered star formation in these pairs. It is
consistent with the results shown in \autoref{Fig3}, \autoref{Fig4}
and \autoref{bias}. The environmental factors dominate after redshift
$2$ due to significant evolution in the local density and
clustering. The enhancement in the local density and clustering
increases the probability of interaction between galaxies. The SFR is
strongly enhanced when two gas-rich galaxies interact with each
other. The starburst activity consumes the available gas in a galaxy,
gradually depleting its cold gas reservoir. The median SFR is reduced
at all pair separations with decreasing redshift after $z=1.49$. The
gradual depletion of the cold gas reservoir causes a reduction in the
median SFR with decreasing redshift.


\begin{figure*}[htbp!]
\centering
\includegraphics[width=14cm]{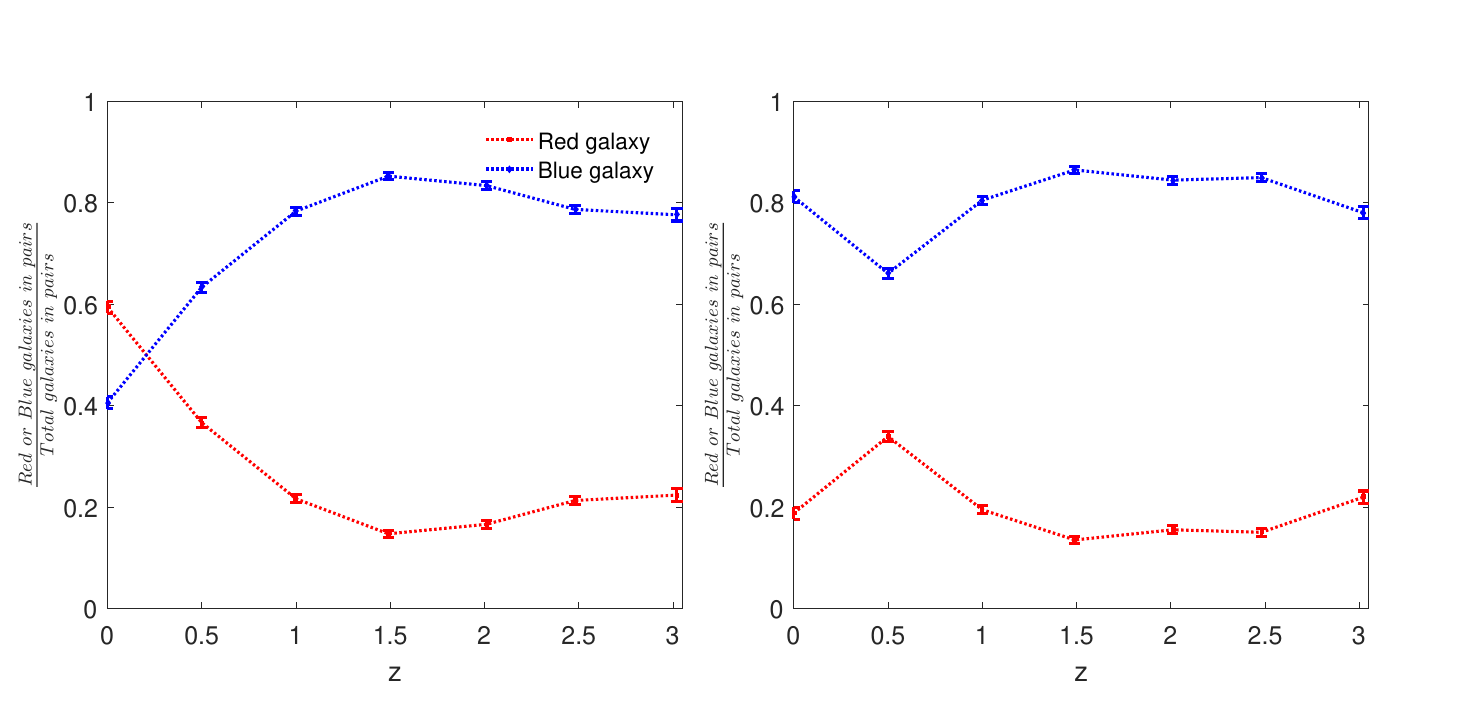}
\caption{The left panel shows the ratio of the red or blue galaxies in
  pairs to the total galaxies in pairs at different redshifts. The
  right panel shows the same but only considering the paired galaxies
  with their second nearest neighbour at a distance $>500$ kpc. We
  consider only the pairs with mass ratio $1 \leq \frac{M_1}{M_2} \leq
  10$ and separation $r \geq 30$ kpc. We show the 1$\sigma$ binomial
  errorbars at each data point.}
\label{interact2}
\end{figure*}

\begin{figure*}[htbp!]
\centering
\includegraphics[width=14 cm]{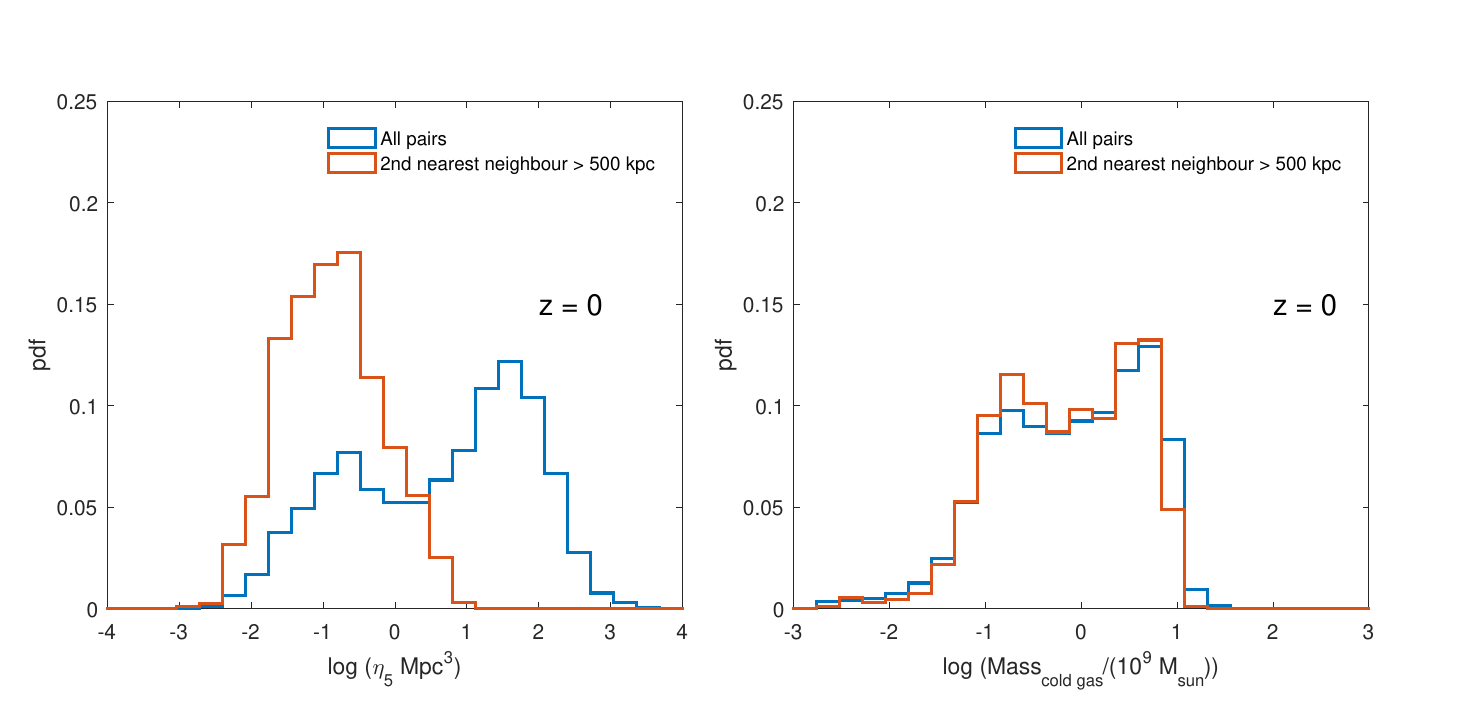}
\caption{The left panel of this figure compares the local density
  distribution of all galaxy pairs with the pairs having their second
  nearest neighbour at a distance $>500$ kpc. The right panel compares
  the cold gas mass distribution for the same pairs.}
\label{Fig7}
\end{figure*}

We calculate the ratio of the cumulative median SFR of the paired
galaxies and the control-matched isolated galaxies at each pair
separation. We show the results in the bottom left panel of
\autoref{interact1}. The results are shown for $4$ different
redshifts. We note that the interaction induced SFR enhancement in
galaxy pairs are more pronounced at $z=0$. In the control-matched
pairs, we observe an enhancement in SFR within the range of
$2-4$. Different studies \citep{ellison08, patton11, shah22} suggest
that interacting galaxies can have SFR enhancements of up to $2-3$
times compared to isolated galaxies. Our results are in close
agreement with these studies. The SFR enhancement decreases with the
increasing pair separation at each redshift. However, the ratio
remains larger than $1$ even at pair separations of $200$ kpc. This
clearly suggests that interactions between galaxies are not driving
the SFR enhancement at such large separation. We have control matched
our pair samples in stellar mass and local density. The large-scale
environment of paired galaxies and isolated galaxies in our sample
could be different. It is known from earlier studies \citep{scudder12,
  cohen17, maret22} that there are significant differences in the star
formation rates between galaxies embedded within different types of
supercluster environments. The signal of enhanced SFR at such large
scales in our study may be influenced by the large-scale environmental
differences between the pair samples and the control samples. Further,
the SFR decreases in all galaxies at lower redshifts due to the
reduced size of their gas reservoirs. The ratio of the median SFR, as
depicted in the bottom left panel of \autoref{interact1}, does not
quantify the actual SFR levels in galaxy pairs. Instead, it
specifically measures the effectiveness of tidal interactions in
promoting SFR enhancement within paired systems when compared to
isolated galaxies. Our analysis suggests that despite the reduced gas
reservoir, interactions are more effective in promoting enhancement of
SFR in the present universe.

We repeat our analysis for the galaxy pairs for which the member
galaxies do not have their second nearest neighbour within $500$
kpc. The results are shown in the top right and bottom right panels of
\autoref{interact1}. We observe a similar trend in these
panels. However, the degree of SFR at smaller pair separation is
higher for these pairs. Interestingly, the SFR enhancement is less
pronounced for these pairs (bottom right panel of \autoref{interact1})
at smaller pair separation. The isolated pairs are less affected by
their surrounding environment. Most of these isolated pairs do not
belong to the denser regions like groups and clusters where a higher
frequency of close interactions can elevate the SFR.

 It would also be interesting to study the fraction of red or blue
 galaxies in paired systems as a function of redshift. We show the
 fraction of red and blue galaxies that are members of galaxy pairs
 having a separation of $<200$ kpc in the left panel of
 \autoref{interact2}. We find that a large fraction ($\sim 75\%$) of
 interacting galaxies are blue at $z\sim 3$. The fraction of blue
 galaxies slowly increases upto $z=1.5$, after which it steadily
 declines to $\sim 40\%$ at $z=0$. The fraction of red galaxies rises
 from $\sim 20\%$ at $z=1.5$ to $\sim 60\%$ at $z=0$, indicating a
 decline in the SFR due to the gradual exhaustion of the cold gas
 supply in galaxies. Both galaxy interactions and environmental
 effects may have a role in such transformation. The fraction of red
 and blue galaxies becomes nearly equal around a redshift of
 $z=0.25$. This crossover could be sensitive to the background
 cosmology and the details of the galaxy formation and evolution. We
 propose that such a transition may be used as a possible diagnostic
 of the different cosmological parameters and the parameters of the
 galaxy formation models.

 We now shift our attention to the isolated pairs for which the second
 nearest neighbour must lie beyond a distance of $500$ kpc. We show
 the fraction of red and blue galaxies in such pairs as a function of
 redshift in the right panel of \autoref{interact2}. The proportion of
 blue galaxies among isolated pairs increases gradually from
 approximately $80\%$ at $z=3$ to about $90\%$ at $z=1.5$, then
 declines until $z=0.5$, following a trend similar to pairs where the
 second nearest neighbour lies within $500$ kpc. Interestingly, the
 proportion of blue galaxies in such pairs begins to increase again at
 $z<0.5$, which is near the redshift when dark energy begins to
 dominate the expansion of the universe. The environments within the
 large-scale structures like galaxy clusters became denser over time
 even after the dark energy domination, affecting the evolution of
 galaxies within them.  Galaxy interactions within denser regions can
 trigger starburst activity and gas loss through various physical
 mechanisms. These processes ultimately deplete the gas reservoir in
 these galaxies, causing them to transition into red systems. A
 smaller percentage ($\sim20\%$) of red galaxies found in isolated
 pairs at redshift $z=0$ suggests that these pairs are less influenced
 by environmental factors, especially post $z=0.5$. The frequency of
 interactions in low-density regions would decrease significantly
 after the universe entered a phase of accelerated expansion.
 According to the equilibrium model \citep{dekel09, bouche10, dave11,
   lilly13}, galaxies maintain a balance between gas inflow, outflow,
 and star formation activities. While interactions and mergers can
 disrupt this equilibrium, galaxies typically return towards stability
 over time. Deviations in SFR from this equilibrium are closely linked
 to the available gas fraction, regulated by inflows and outflows of
 gas. In less dense environments, where environmental impacts and gas
 outflows become less severe, isolated galaxy pairs can either
 maintain or even increase their SFR compared to counterparts in
 denser environments, where star formation is more likely to be
 suppressed.

In the left panel of \autoref{Fig7}, we compare the local density
distributions of all pairs with the pairs having their second nearest
neighbour beyond $500$ kpc. The pairs with their second nearest
neighbour at a distance $>500$ kpc reside in the less dense
environments. The right panel of \autoref{Fig7} shows that these pairs
have a relatively larger cold gas reservoir. We apply a
Kolmogorov-Smirnov test to compare the PDFs of the two types of pairs
and find that the null hypothesis can be rejected at $>99\%$
confidence in each case. A larger cold gas supply would delay the
quenching in such galaxy pairs. The differences in the size of the gas
reservoirs may arise due to their different evolutionary history.

\section{Conclusions}

We classify the red and blue galaxies in the EAGLE simulation at
different redshifts and study the evolution of their environment,
clustering and physical properties as a function of redshift. We also
study the cumulative median SFR and the SFR enhancement in interacting
galaxies as a function of the pair separation at different
redshifts. These together allow us to study the combined roles of
environment and galaxy interactions in the evolution of galaxies. Our
primary results are as follows:

(i) In the present universe, the red galaxies preferentially reside in
the denser environments, whereas the blue galaxies inhabit the
lower-density regions. The red galaxies are more strongly clustered
than the blue galaxies. The blue galaxies have a larger cold gas mass
and higher sSFR than the red galaxies. The red galaxies dominate the
blue galaxies at the lower and higher ends of the stellar mass
distribution. The lower and higher stellar mass populations in the red
sequence could be the quenched satellites and the massive central
galaxies, respectively.

(ii) The differences in the local density and the clustering strength
of the red and blue galaxies rise strongly with the decreasing
redshift after $z=2$. The relative bias between the red and the blue
galaxies increases from $1$ to $2.5$ between $z=3-0$. The asymmetry in
the cold gas mass and sSFR distributions of the red and blue galaxies
are amplified by the environmental factors after $z=2$.

(iii) The dichotomy between the star formation history of the galaxies
exists at all redshifts between $z=0-3$. The cold gas mass and the
sSFR distributions of the red and blue galaxies differ significantly
between $z=3-2$ even though the differences in environments and
clustering strengths are relatively smaller during this period. It
indicates that the dichotomy between the two populations does not
arise due to the environment alone. Such differences may be related to
the assembly history and the gas accretion efficiency of their host
dark matter halos.

(iv) Interacting galaxy pairs at a given separation have a larger
median SFR at higher redshifts. This could be related to the
increasing gas fractions at higher redshift. The median SFR at all
separations reduce with the decreasing redshift, implying a gradual
exhaustion of the cold gas supply. An anomalous increase in the median
SFR at smaller pair separation is observed around $z\sim 2$. It
indicates that environmental factors start to dominate around this
redshift, increasing the frequency of interactions and leading to
starbursts in close pairs.

(v) The median SFR of the interacting pairs are elevated compared to
the mass-matched and environment-matched isolated galaxies at nearly
all pair separations within $r<200$ kpc. The enhancement signal
diminishes with increasing pair separation suggesting that
interactions are less influential at larger separations. The SFR
enhancement at larger separations implies potential large-scale
environmental differences between galaxies in pairs and isolated
galaxies. Despite diminished gas reservoirs at lower redshifts,
interactions effectively boost star formation rates in the current
epoch.

(vi) The fraction of blue galaxies in interacting pairs increases with
the decreasing redshift upto $z=1.5$. It steeply declines after this
redshift until the present. It may arise due to the depletion of cold
gas supply in the interacting galaxies due to starbursts and other
environmental effects. Until $z=0.5$, the blue fraction declines in a
similar way in the pairs that have their second nearest neighbour at a
distance $>500$ kpc. However, the blue fraction begins to increase in
isolated pairs at $z<0.5$, possibly because the frequency of
interactions and gas loss decreased significantly in the low density
regions after the universe entered a phase of accelerated
expansion. The isolated galaxy pairs in low-density environments have
a larger cold gas supply, which enables them to maintain star
formation activity and delay the quenching.

We compare our results with the other studies with the EAGLE and
Illustris TNG simulations \citep{nelson19}. \cite{trayford16} study
the evolution of galaxy colour using the EAGLE simulation. They find
that the red sequence builds up at $z\sim 1$ due to the quenching of
low-mass satellite galaxies and AGN feedback in the more massive
central galaxies. \cite{wright19} show that the quenching time scales
for both the central and satellite galaxies in the EAGLE simulation
decrease with the increasing stellar mass. They find a growth in the
fraction of passive galaxies at lower redshifts. Using TNG100 and
TNG300, \cite{donnari21} show that $70\%-90\%$ of centrals and
satellites with mass $\geq 10^{10} M_{\odot}$ are predicted to be
quenched between $0\leq z \leq 0.5$. Another study \citep{walters22}
with the TNG simulation finds that the slow-quenching galaxies are
about twice as common as the fast-quenching galaxies in the redshift
range $0.7<z<2$. Our results are consistent with these findings.

Our results are also consistent with the observational findings
\citep{madau96, tran10, forster20, gupta20} indicating that the
hydrodynamical simulations are a powerful tool for studying galaxy
formation and evolution. A recent study \citep{cerdocino24} with the
photometric data from the S-PLUS DR4 survey find a high fraction of
red galaxies in paired systems. They show that the red pair fractions
increase in closer pairs and pairs of similar luminosity, indicating
that shared environments and interactions have important roles in the
evolution of galaxy colour. Our analysis with the EAGLE simulation
also indicates that both interactions and environment play crucial
roles in the galaxy evolution. However, the dichotomy between the red
and blue galaxies may not arise due to the environment alone. The
differences in the evolutionary history may lead to galaxies with
different cold gas content and star formation rates. The environment
and interactions at lower redshifts amplify these differences,
splitting the galaxies into two distinct populations.

\section*{ACKNOWLEDGEMENT}
We sincerely thank an anonymous reviewer whose insightful
comments and suggestions helped us to significantly improve the
draft. BP would like to acknowledge financial support from the SERB,
DST, Government of India through the project CRG/2019/001110. BP would
also like to acknowledge IUCAA, Pune for providing support through
associateship programme.

The authors acknowledge the Virgo Consortium for making their
simulation data publicly available. The EAGLE simulations were
performed using the DiRAC-2 facility at Durham, managed by the ICC,
and the PRACE facility Curie based in France at TGCC, CEA,
Bruy\`{e}res-le-Ch\^{a}tel.

\end{document}